\documentclass[conference]{IEEEtran}
\IEEEoverridecommandlockouts
\usepackage{cite}
\usepackage{amsmath,amssymb,amsfonts}
\usepackage{algorithmic}
\usepackage{graphicx}
\usepackage{textcomp}
\usepackage{xcolor}
\usepackage{siunitx}
\usepackage{tabularx}
\usepackage{booktabs}
\usepackage{cite}
\def\BibTeX{{\rm B\kern-.05em{\sc i\kern-.025em b}\kern-.08em
    T\kern-.1667em\lower.7ex\hbox{E}\kern-.125emX}}

\begin{document}

\title{J-Net: Improved U-Net for Terahertz Image Super-Resolution}
\author{
    \IEEEauthorblockN{Woon-Ha Yeo}
    \IEEEauthorblockA{\textit{Department of Artificial Intelligence Convergence} \\
    \textit{Sahmyook University}\\
    Seoul, Republic of Korea \\
    canal@syuin.ac.kr}
\and
    \IEEEauthorblockN{Seung-Hwan Jung}
    \IEEEauthorblockA{\textit{Department of Artificial Intelligence Convergence} \\
    \textit{Sahmyook University}\\
    Seoul, Republic of Korea \\
    jshwan0828@syuin.ac.kr}
\and
    \IEEEauthorblockN{Seung Jae Oh}
    \IEEEauthorblockA{\textit{YUHS-KRIBB Medical Convergence Research Institute} \\
    \textit{Yonsei University College of Medicine}\\
    Seoul, Republic of Korea}
\and
    \IEEEauthorblockN{Inhee Maeng}
    \IEEEauthorblockA{\textit{YUHS-KRIBB Medical Convergence Research Institute} \\
    \textit{Yonsei University College of Medicine}\\
    Seoul, Republic of Korea}
\and
    \IEEEauthorblockN{Eui Su Lee\IEEEauthorrefmark{1}\thanks{\IEEEauthorrefmark{1}Corresponding author.}}
    \IEEEauthorblockA{\textit{Electronics and Telecommunications Research Institute (ETRI)}\\
    Daejeon, Republic of Korea \\
    euisu@etri.re.kr}
\and
    \IEEEauthorblockN{Han-Cheol Ryu\IEEEauthorrefmark{1}}
    \IEEEauthorblockA{\textit{Department of Artificial Intelligence Convergence} \\
    \textit{Sahmyook University}\\
    Seoul, Republic of Korea \\
    hcryu@syu.ac.kr}
}

\maketitle

\begin{abstract}
Terahertz (THz) waves are electromagnetic waves in the 0.1 to 10 THz frequency range, and THz imaging is utilized in a range of applications, including security inspections, biomedical fields, and the non-destructive examination of materials. However, THz images have low resolution due to the long wavelength of THz waves. Therefore, improving the resolution of THz images is one of the current hot research topics.
We propose a novel network architecture called J-Net which is improved version of U-Net to solve the THz image super-resolution. It employs the simple baseline blocks which can extract low resolution (LR) image features and learn the mapping of LR images to high-resolution (HR) images efficiently.
All training was conducted using the DIV2K+Flickr2K dataset, and we employed the peak signal-to-noise ratio (PSNR) for quantitative comparison. In our comparisons with other THz image super-resolution methods, J-Net achieved a PSNR of 32.52 dB, surpassing other techniques by more than 1 dB. J-Net also demonstrates superior performance on real THz images compared to other methods.
Experiments show that the proposed J-Net achieves better PSNR and visual improvement compared with other THz image super-resolution methods.
\end{abstract}


\section{Introduction}
Terahertz (THz) imaging, operating within the 0.1 to 10 THz frequency range, is a rapidly evolving field with applications spanning from security inspections to biomedical and materials science. The unique characteristics of THz waves, bridging the gap between microwaves and infrared radiation, offer distinct advantages in these applications. Despite its potential, THz imaging is hindered by inherent limitations in resolution due to the longer wavelengths of THz waves compared to visible light. This results in images that are typically low in resolution, blurred, and noisy, making detailed analysis challenging.

The quest to enhance the resolution of THz images has led to a bifurcated approach: enhancing the imaging hardware and adopting super-resolution techniques. While hardware advancements can yield improvements, they often come with increased costs and complexity. Super-resolution methods, on the other hand, provide a more economical and flexible alternative, leveraging existing imaging systems to achieve high-resolution (HR) image reconstruction. The traditional methods, including deconvolution techniques like the Lucy–Richardson algorithm and various interpolation methods, have been instrumental in initial improvements. However, these approaches often fall short in recovering high-frequency details and handling noise variations.

The advent of deep learning and convolutional neural networks (CNNs) has revolutionized the field of image super-resolution. Pioneering works like the SRCNN~\cite{dongSRCNN2014}, and subsequent advancements like the VDSR~\cite{kim2016vdsr} and ESPCN~\cite{shi2016espcn} networks, have demonstrated superior performance in feature extraction and nonlinear modeling, significantly outperforming traditional methods. Numerous studies in THz imaging frequently adopt models originally proposed for optical image super-resolution in their research~\cite{Long_Wang_You_Yang_Wang_Liu_2019, Ruan_Tan_Chen_Wan_Cao_2022}. However, the direct application of these networks to THz imaging has been limited due to the unique degradation model of THz images, which includes blurring, downsampling, and noising with spatially variable characteristics.

In light of these challenges, our work introduces a novel network architecture, J-Net, an enhanced version of the U-Net framework~\cite{ronneberger2015unet}, which has demonstrated exceptional performance in the field of image restoration. Consequently, it is considered to be effectively applicable for image super-resolution as well, specifically tailored for THz image super-resolution. J-Net is designed to efficiently extract features from low-resolution THz images and learn the mapping to HR images. Unlike previous methods, J-Net is equipped to handle the unique degradation models of THz imaging, incorporating elements that address blurring, noise, and spatial variability. Through rigorous experimentation and comparison with established super-resolution methods, including the Lucy–Richardson deconvolution, Long et al.~\cite{Long_Wang_You_Yang_Wang_Liu_2019} and Ruan et al.~\cite{Ruan_Tan_Chen_Wan_Cao_2022}, J-Net demonstrates superior accuracy and visual improvement in THz image super-resolution. This paper presents a comprehensive analysis of J-Net's performance, showcasing its effectiveness through quantitative metrics such as peak signal-to-noise ratio (PSNR) on the DIV2K+Flickr2K dataset, along with qualitative assessments on real THz images.

\section{Proposed Method}
In this section, we begin by introducing the terahertz imaging system employed in this study, followed by an overview of the degradation model used to construct our training dataset. Subsequently, we provide a detailed explanation of the overall architecture of the proposed J-Net.

\subsection{Terahertz Imaging System}
\begin{figure}[htbp]
\includegraphics[width=\columnwidth]{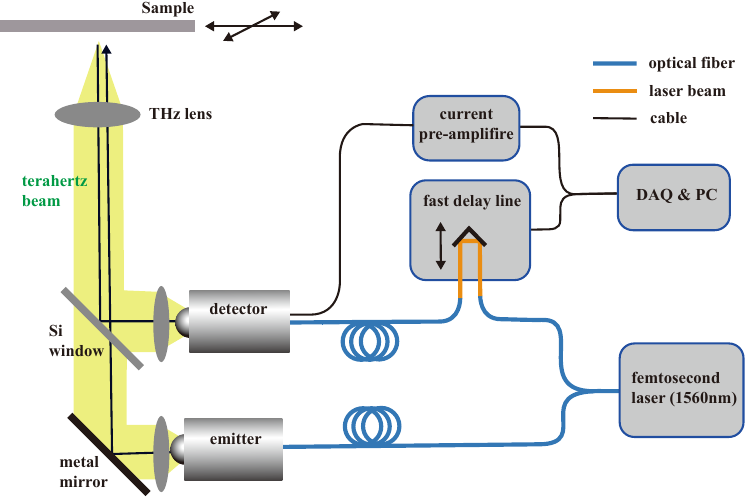}
\caption{THz imaging system.\label{system}}
\end{figure}

We used a reflection mode THz imaging system to acquire images and THz waveforms as shown in Figure~\ref{system}. The THz pulse was obtained using a femtosecond laser and a photoconductive antenna. The femtosecond laser had a central wavelength of \SI{1.5}{\micro\metre} and a pulse width of 80 fs. A fiber-coupled antenna (TERA15-TX-FC, Menlo Systems) was used as a photoconductive antenna for THz pulse generation, and another fiber-coupled dipole antenna (TERA15-RX-FC, Menlo Systems) was used to detect the THz signal. To rapidly acquire the THz signal, we used an ultra-fast scanner with a frequency and time-width setting of 20 Hz and 30 ps, respectively.

The THz signal was amplified through a low-noise current preamplifier (SRS570, Stanford Research) and then digitized via a data acquisition board. The generated THz pulses were guided through a polymethylpentene (TPX) lens and plate mirrors. The TPX lens focused the THz signal onto the sample stage, and the signal reflected from the sample was guided through a silicon beam splitter, another TPX lens, and into the detector. In particular, all components of the system, except the sample, were placed within a dry air chamber to avoid signal distortion due to water vapor absorption. The sample was placed in a 3 mm thick crystallized z-quartz window (subject to change). The time-domain signal obtained at each pixel was converted to the frequency domain via Fast-Fourier Transform (FFT), thereby endowing each pixel with broadband THz wave characteristics. The sample was moved along the x and y axes to form a 2D image based on the response of the THz pulse signal. Each pixel's time-domain signal can be transformed into the frequency domain through FFT, giving each pixel in the image its unique broadband THz signal characteristics.

\subsection{Degradation Model}
To train a model for reconstructing low-resolution (LR) images into high-resolution (HR) images, pairs of HR and LR images are needed. However, obtaining these pairs from THz images is challenging. Therefore, it is necessary to degrade HR images to create LR images that closely resemble the characteristics of THz images. This enables the training of the model using these artificially created pairs. The degradation model used in this process is of great importance. It must be effectively designed to simulate the real-world behavior of THz images to ensure the model's performance and accuracy in practical applications.

In the context of image super-resolution, the degradation model is typically represented by the following equation:
\begin{equation}
    I_{LR} = (I_{HR} \ast k) \downarrow_s + n
\end{equation}
where \( I_{LR} \) denotes the LR image, \( I_{HR} \) represents the HR image, \( k \) is the blurring kernel, \( \downarrow_s \) signifies down-sampling by a scaling factor, and \( n \) is the additive noise. This model describes how a high-resolution image is transformed into its low-resolution counterpart. It involves the application of a blurring kernel \( k \) to the high-resolution image, followed by down-sampling with a scaling factor \( s \), and finally the addition of noise \( n \). The goal of THz image super-resolution is to restore the $I_{HR}$ from the $I_{LR}$.

The Gaussian blur kernel is often treated as the point spread function (PSF) of the THz imaging systems. The PSF depends on the imaging system and usually can be approximately an isotropic Gaussian blur kernel~\cite{Long_Wang_You_Yang_Wang_Liu_2019}. Given the presence of defocusing in real-world imaging and the use of a frequency-seeping source by the system, the blur kernel undergoes changes within a specific range. As a result, to enhance the model's robustness, all conceivable blur kernels should be included in the training set. Because obtaining all the blur kernels experimentally is challenging, and THz beams usually follow the Gaussian distribution~\cite{van1990characterization, jepsen1996generation}, we substitute the actual PSF with the Gaussian blur kernels. The equation is as follows:
\begin{equation}
    G(x,y) = \frac{1}{2\pi\sigma^2}\exp{[-\frac{(x^2+y^2)}{2\sigma^2}]},
\end{equation}
where $\sigma$ represents the standard deviation, which is the width of the Gaussian kernel. It is important to include multiple levels of noise in the training set, as THz waves may produce varying levels of noise when interacting with different substances.

When forming a Gaussian blur kernel, the standard deviation \( \sigma \) is randomly selected from a range between $\alpha$ and $\beta$. This can be mathematically represented as:
\begin{equation}
    \sigma \sim \mathcal{U}(\alpha, \beta)
\end{equation}
where \( \mathcal{U}(\alpha, \beta) \) denotes the uniform distribution in the range. The random selection of \( \sigma \) introduces variability in the blurring process, which is a crucial aspect in simulating different levels of image degradation.
 
\subsection{Network Architecture}
\begin{figure}[htbp]
\includegraphics[width=\columnwidth]{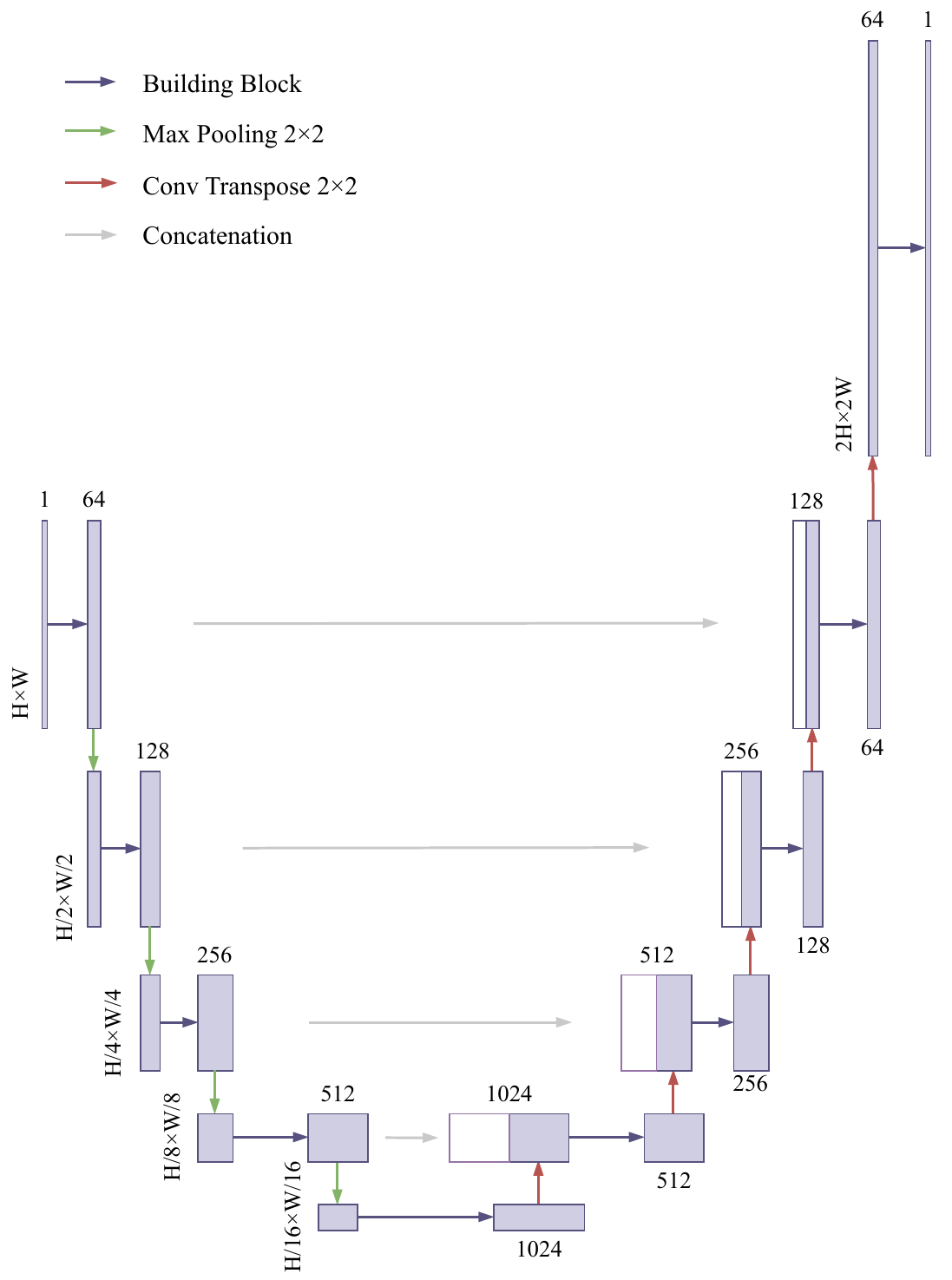}
\caption{Network architecture of the proposed J-Net.\label{jnet}}
\end{figure}

In this section, we build a novel architecture for THz image super-resolution. We modify the U-Net~\cite{ronneberger2015unet} architecture by incorporating an additional expansive path at the end. The modification is aimed at increasing the size of the feature map for super-resolution. This results in a shape that resembles the letter "J" rather than "U" as shown in Figure~\ref{jnet}. The original U-Net architecture is a convolutional neural network (CNN) often used for image restoration. Its structure, shaped like the letter "U," comprises two parts: the contraction path for downsampling and the expansive path for upsampling.

\begin{enumerate}
    \item Contraction Path: This section captures contextual information from the image, utilizing convolutional layers followed by max pooling. This process reduces the image's spatial dimensions, enabling the network to extract essential features, which is crucial for restoring degraded parts of an image.
    \item Expansive Path: In this part, the network upscales the feature maps to reconstruct the image at its original resolution. The use of transposed convolutions or up-convolution layers is common here. This path also involves the concatenation of feature maps from the contraction path, allowing the network to utilize both high-level and detailed information, which is critical for accurately restoring image details.
    \item Skip Connections: A standout feature of U-Net is its use of skip connections. These connections help transfer detailed information by linking feature maps from the contraction path directly to the expansive path. This feature is particularly beneficial in image restoration as it allows for the preservation and incorporation of fine details in the restored image.
\end{enumerate}

The key differences from the basic U-Net in our approach lie in the addition of an extra expansive path towards the end and the building blocks used. We adopted the building block as the simple baseline block~\cite{chen2022simple}. The details of the block are described in the following subsection.

\subsection{Building Block}
\begin{figure}[htbp]
\includegraphics[width=\columnwidth]{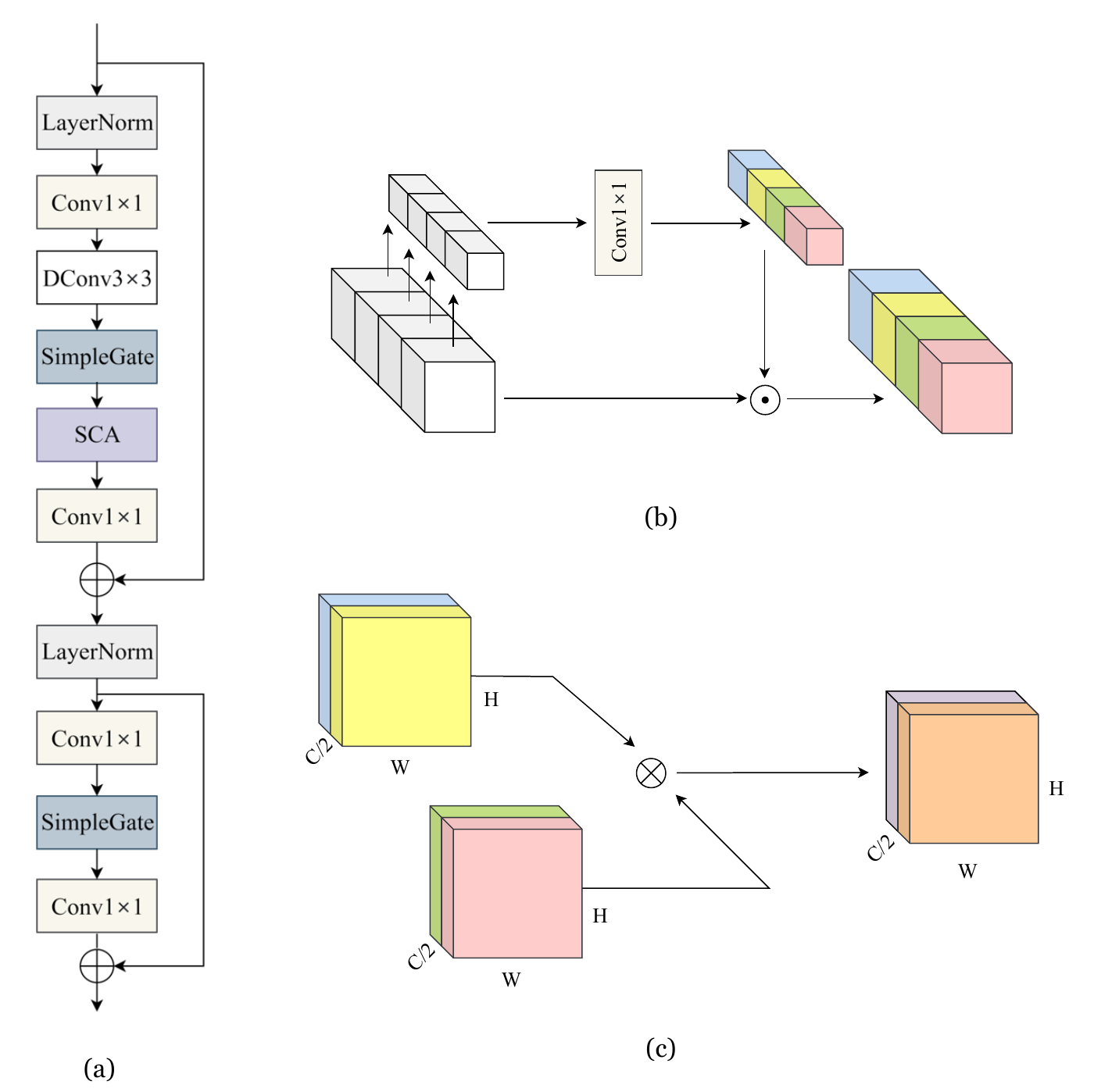}
\caption{Description of building block used in J-Net. (a) Baseline block used in J-Net; (b) Simplified channel attention (SCA) module; (c) Simple gate. \label{block}}
\end{figure}

The J-Net framework utilizes a simple baseline block~\cite{chen2022simple} that extends beyond mere convolution layers. This essential block combines innovative components used in the latest image restoration methods. First, it incorporates Layer Normalization (LN)~\cite{ba2016layer} which not only streamlines the block's configuration but also stabilizes the training phase. This stability allows for an increased learning rate, boosting it from 0.0001 to 0.001, enhancing both deblurring and noise reduction capabilities~\cite{chen2022simple}. Second, channel attention brings computational efficiency and global information to the feature maps. Third, nonlinear activation functions such as GELU~\cite{hendrycks2016gaussian} are replaced by a simple gate which is an element-wise product of feature maps. The simple gate can produce the effect of the nonlinear activation function and leads to performance gain. Finally, the block integrates LN, convolution, simple gate, and simplified channel attention (SCA). It delivers commendable results in image restoration. Its combination of simplicity and effectiveness is noteworthy. This block and its components are described in Figure~\ref{block}.

\section{Experimental Results}
\subsection{Datasets}
Our research utilized a combination of the DIV2K~\cite{agustsson2017div2k} and Flickr2K~\cite{lim2017flickr2k} datasets for training the THz image super-resolution model. These datasets, integral in the image processing domain, provided a rich variety of scenarios. DIV2K, with its 1,000 diverse images, including 800 for training and 200 for testing and validation, offered high-quality 2K resolution images. Flickr2K complemented this with additional versatility, allowing for the creation of datasets from training or validation sets and encompassing both low and high-resolution images. This comprehensive approach using both datasets was crucial for effectively training our THz image super-resolution model.

\begin{figure}[htbp]
\includegraphics[width=\columnwidth]{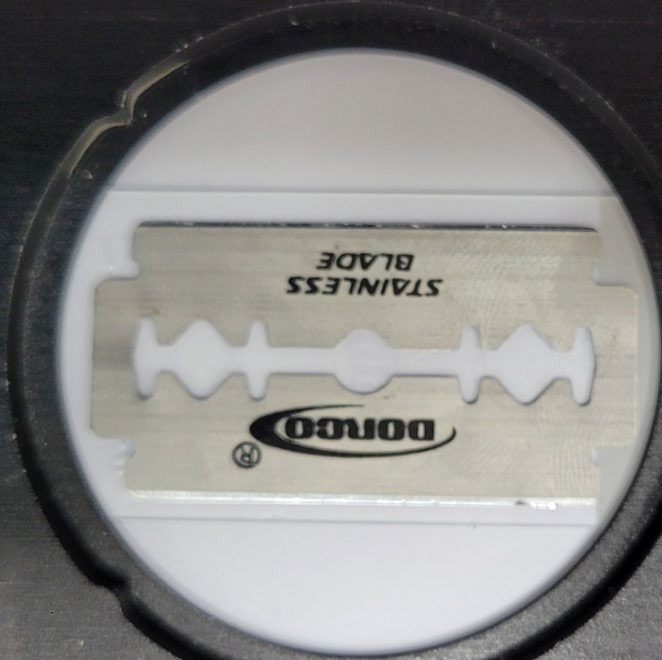}
\caption{Optical image of a metal knife used for measuring THz image.\label{knife}}
\end{figure}

To test the proposed method, we measure a metal knife (shown in Figure~\ref{knife}) using THz time domain spectroscopy (TDS) system.

\subsection{Implementation Details}
In the training process, data augmentation was applied to DIV2K~\cite{agustsson2017div2k} and Flickr2K~\cite{lim2017flickr2k} datasets using random flip and rotation with the probability of 0.5. The number of the baselines blocks is all set to 2, and the width in all the blocks was set to 64. The mini-batch size was 32, and the patch size was 96$\times$96 (HR images are randomly cropped to 96$\times$96). The initial learning rate was set to 1e-3 and gradually reduced to 1e-6 with the cosine annealing schedule~\cite{loshchilov2016sgdr}. Our model was trained for up to 200K iterations. We implemented the code using PyTorch deep learning framework~\cite{paszke2017automatic}. All models were conducted by eight NVIDIA A100 GPUs. We used Adam~\cite{kingma2014adam} optimizer with $\beta_1=0.9$, $\beta_2=0.99$, weight decay 1e-4 to optimize the following loss function:
\begin{equation}
    \mathcal{L}_{MSE} = \frac{1}{N} \sum_{i=1}^n \vert\vert I_{HR}-I_{SR}\vert\vert^2,
\end{equation}
where $N$ denotes the number of training samples, $I_HR$ denotes the HR image, and $I_SR$ represents predicted image by a trained super-resolution (SR) model. $L_{MSE}$ denotes mean squared error between the HR images and the predicted SR images.

\subsection{Results and Discussion}
\subsubsection{U-Net vs. J-Net}
To show the effectiveness of the proposed J-Net, we have compared the original U-Net~\cite{ronneberger2015unet} and the J-Net. In this section, naive convolution layers are used as the building blocks of the network. The U-Net is originally designed for medical image semantic segmentation; however, it also achieves great performance on image restoration. The U-shaped network structure is then widely used in the image restoration field~\cite{wang2022uformer,zamir2021multi,chen2022simple,zamir2020learning,zamir2022restormer,zhang2021plug}. Since image super-resolution is one of image restoration, U-Net also could be effective for image super-resolution. We conducted experiment the structure of network architecture for three types: the series of basic convolution layers with following a PixelShuffle (PS) layer~\cite{shi2016espcn} called Flat U-Net, the U-Net with the PS layer, and the J-Net. We measured the peak signal-to-noise ratio (PSNR) on DIV2K validation set. Table~\ref{unetvsjnet} shows the PSNR results on the DIV2K validation set, and indicates that the J-Net is more efficient than other structures in image super-resolution.

\begin{table}[htbp] 
\caption{U-Net vs. J-Net\label{unetvsjnet}}
\newcolumntype{C}{>{\centering\arraybackslash}X}
\begin{tabularx}{\columnwidth}{CCCC}
\toprule
\textbf{} & Flat U-Net & U-Net	& J-Net\\
\midrule
PSNR	& 30.17	& 31.38	& \textbf{31.53}	\\ 
\bottomrule
\end{tabularx}
\end{table}

\subsubsection{Variation of Degradation Parameter}
\begin{figure*}[htbp]
\centering
\includegraphics[width=0.9\textwidth]{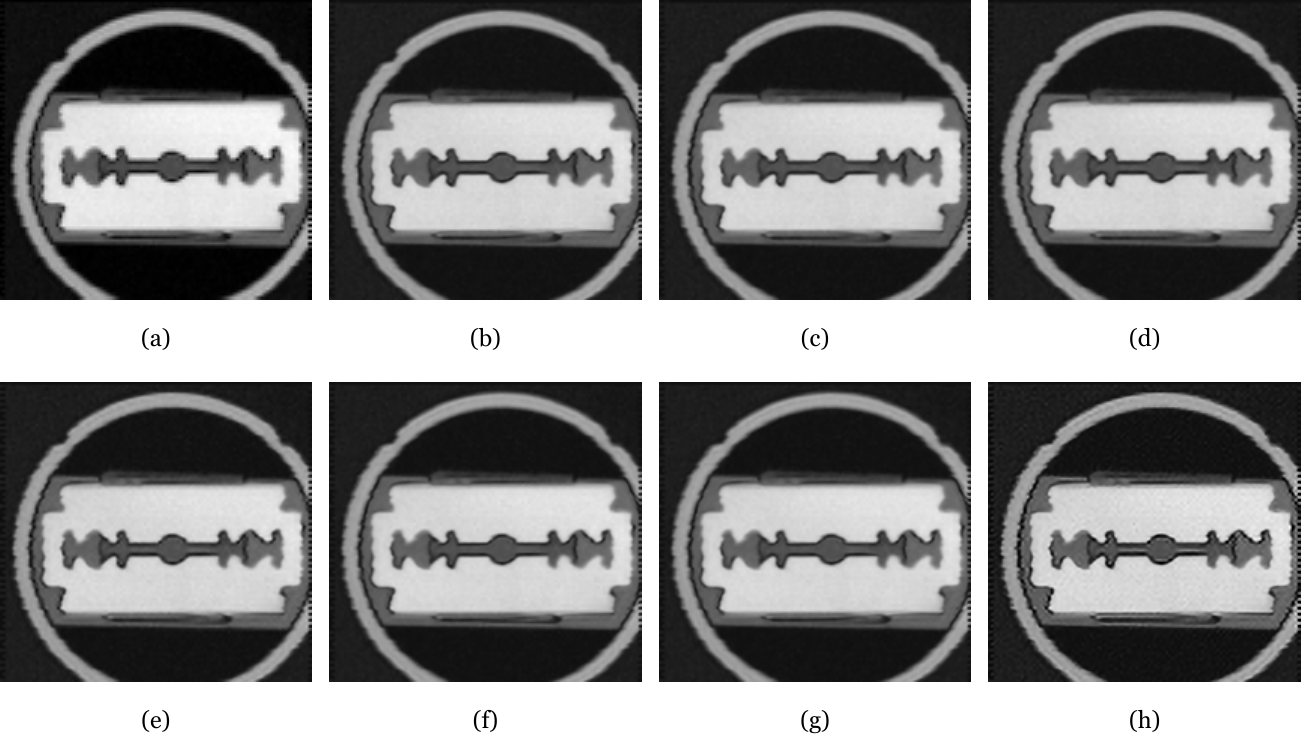}
\caption{Experimental results on real THz image of metal knife. (a) Original THz image obtained at a frequency of 1.0 THz; (b) $\alpha=0,\beta=1$; (c) $\alpha=0,\beta=1$; (d) $\alpha=0,\beta=5$; (e) $\alpha=0,\beta=10$; (f) $\alpha=0.1,\beta=1$; (g) $\alpha=0.1,\beta=3$; (h) $\alpha=1,\beta=2$.\label{stdvalue}}
\end{figure*}

We randomized the standard deviation $\sigma$ value of the Gaussian blur between $\alpha$ and $\beta$. We show qualitative and quantitative results for different values of $\alpha$ and $\beta$. In Table~\ref{result_sigma}, the smaller the difference between alpha and beta, the higher the PSNR value, and if the difference is the same, the PSNR is higher for higher alpha-beta values. You can see that the PSNR is significantly higher when alpha is 0.1 than when it is 0.
Figure~\ref{stdvalue} shows the inference results using the model in the table. (a) is the original and the rest are the results of doubling the image resolution compared to the original image. Note the smoother boundaries compared to the original. It can be seen that the results for $\alpha = 0$ all tend to follow the noise of the original image, while for 
$\alpha = 0.1$ the noise is smoothed out. At an alpha of 1, the results don't seem to fully reflect the PSF of the THz system. Therefore, it is important to choose the appropriate values of $\alpha$ and $\beta$.
\begin{table}[htbp] 
\caption{Comparison of performance based on the value of $\alpha$ and $\beta$ in the DIV2K validation set.\label{result_sigma}}
\newcolumntype{C}{>{\centering\arraybackslash}X}
\begin{tabularx}{\columnwidth}{CCC}
\toprule
\textbf{$\alpha$} & \textbf{$\beta$} & PSNR \\
\midrule
0	& 1  & 34.96 \\ 
0   & 3  & 34.41 \\
0   & 5  & 33.13  \\
0   & 10 & 30.09 \\
0.1 & 1	 & 35.44 \\
0.1 & 3	 & 35.32 \\
1   & 2	 & 35.08 \\
\bottomrule
\end{tabularx}
\end{table}

\subsubsection{Model Comparison}
\begin{figure*}[htbp]
\centering
\includegraphics[width=0.8\textwidth]{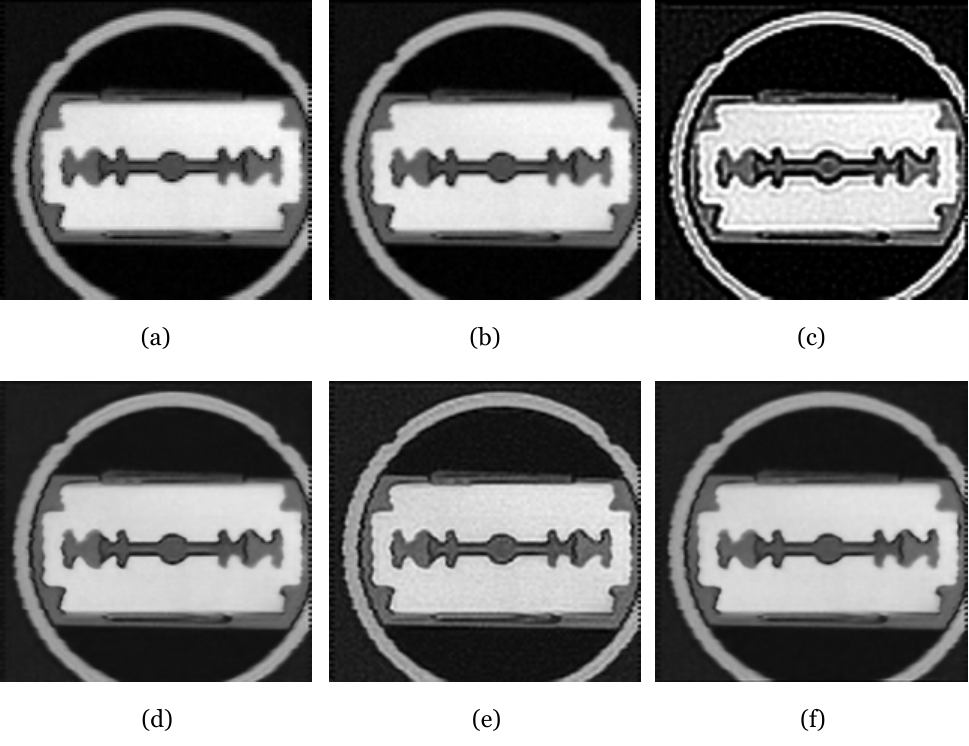}
\caption{Experimental results on real THz image of metal knife. (a) Original THz image obtained at a frequency of 1.0 THz; (b) Bicubic interpolation; (c) Lucy–Richardson deconvolution; (d) Long et al.~\cite{Long_Wang_You_Yang_Wang_Liu_2019}; (e) Ruan et al.~\cite{Ruan_Tan_Chen_Wan_Cao_2022}; (f) J-Net.\label{fig:model}}
\end{figure*}

In this subsection, we compare the performance of the proposed method, bicubic interpolation algorithm, the widely used Lucy–Richardson algorithm, Long et al.~\cite{Long_Wang_You_Yang_Wang_Liu_2019}, and the U-Net on the real world THz image. Since the Lucy-Richardson algorithm is a deconvolution algorithm which cannot improve the resolution of images, we use bicubic interpolation to interpolate the image after applying the Lucy-Richardson algorithm. As can be seen from Table~\ref{tab:model}, the proposed J-Net shows the best PSNR among the deep learning based methods~\cite{Long_Wang_You_Yang_Wang_Liu_2019, Ruan_Tan_Chen_Wan_Cao_2022}. Figure~\ref{fig:model} shows the visual results of different methods on the real THz image of a metal knife. One can see that the proposed method can recover image sharpness well, and the Lucy-Richardson algorithm shows distorted artifact as a result.

\begin{table}[htbp] 
\caption{Comparison of PSNR performance on other THz image super-resolution methods.\label{tab:model}}
\newcolumntype{C}{>{\centering\arraybackslash}X}
\begin{tabularx}{\columnwidth}{CCCC}
\toprule
\textbf{} & Long et al.~\cite{Long_Wang_You_Yang_Wang_Liu_2019} & Ruan et al.~\cite{Ruan_Tan_Chen_Wan_Cao_2022}	& J-Net\\
\midrule
PSNR	& 32.42 & 30.58	& \textbf{32.52}	\\ 
\bottomrule
\end{tabularx}
\end{table}

\section{Conclusions}
Our study introduces J-Net, a new neural network architecture tailored for enhancing the resolution of THz images. In comparison with traditional methods and other deep learning approaches, J-Net shows superior performance in improving image clarity, as evidenced by its higher peak signal-to-noise ratio (PSNR) in tests using the DIV2K+Flickr2K dataset and real-world THz images. The success of J-Net demonstrates its potential to significantly improve the quality of THz imaging in various applications, including security, medical imaging, and materials analysis, making it a noteworthy advancement in the field of image super-resolution.

\section*{Acknowledgment}
This work was supported by Basic Science Research Program through the National Research Foundation (NRF) funded by the Ministry of Science and ICT (grant no. NRF-2021R1F1A1059493) and an Electronics and Telecommunications Research Institute (ETRI) grant funded by the Korea government (Grant No. 23ZB1130)

\bibliography{references}
\bibliographystyle{plain}

\end{document}